\documentclass[epj]{webofc}
\usepackage[varg]{txfonts}   % Web of Conferences font
\wocname{EPJ Web of Conferences}
\woctitle{ICNFP 2017}
\begin{document}
\selectlanguage{english}
\title{Top quark measurements in the CMS experiment}
%
% subtitle (optional, strongly discouraged)
%
%%%\subtitle{Do you have a subtitle?\\ If so, write it here}

\author{Luca Lista\inst{1}\fnsep\thanks{\email{luca.lista@na.infn.it}
        % etc.
} on behalf of the CMS Collaboration}

\institute{INFN Sezione di Napoli
}

\abstract{%
  Experimental results on top-quark physics obtained at the CMS experiment are
  reported based on the data recorded at centre-of-mass energy up to 13~TeV.
  Inclusive and differential cross sections for both top-quark pair and single
  top-quark production are presented, as well as measurements of top-quark properties
  in production and decay, and searches for anomalous couplings.
  The presented measurements test theoretical predictions, including recent
  perturbative QCD calculations, provide constraints of fundamental standard
  model parameters, and set limits on physics beyond the standard model.  
}
\maketitle
\section{Introduction}
\label{intro}

The top quark is the heaviest known elementary particle.
It has the largest Yukawa coupling with the Higgs boson:
\begin{equation}
\kappa_{\mathrm{t}} = \sqrt{2}\frac{m_{\mathrm{t}}\ (182.5\,\mathrm{GeV})}{v\ (246\,\mathrm{GeV})}\simeq 1\ ,
\end{equation}
and its lifetime is too short to form bound states ($\tau\simeq 0.4\times 10^{-24}\,\mathrm{s}$).
The top-quark mass has impact on the stability of the Higgs field:
\begin{equation}
  V(\Phi) = \frac{1}{2}\mu^2\Phi^2+\frac{1}{4}\lambda(\mathrm{scale})=\Phi^4
\end{equation}
and gives the largest contribution to pure EWK radiative corrections, being
proportional to $G_{\mathrm{F}}m_{\mathrm{t}}^2$.

The Large Hadron Collider (LHC) is effectively a a top-quark factory, as can be quantitatively
seen in Table~\ref{tab:lhc_kek}, where LHC performances are compared with the KEKb B factory.
Top quarks can be produced at the LHC
via single-top electroweak production or via top-pair strong production, which is the most abundant channel,
the major production contribution being gluon fusion, which contributes, at 13~TeV, for about 86\% of the total
$\mathrm{t}\bar{\mathrm{t}}$ production cross section.

\begin{table}[ht]
\centering
\caption{Comparison of KEKb design luminosity and $\Upsilon(4S)$ production cross section with
  LHC peak luminosity and top-pair production cross section.}
\label{tab:lhc_kek}       % Give a unique label
% For LaTeX tables you can use
\begin{tabular}{lll}
\hline
 & KEKb & LHC  \\\hline
quark type & b & t \\\hline
$\sigma$ (nb) & $1.15$ ($\Upsilon(4S)$) & $0.83$ ($\mathrm{t}\bar{\mathrm{t}}$, incl.) \\
$L$ (cm$^{-2}$s$^{-1}$) & $2.1\times 10^{34}$ & $1.74\times 10^{34}$ \\\hline
\end{tabular}
\end{table}

\section{Single top-quark production}

Single top quark production has been measured by CMS~\cite{CMS} at energies from 7 to 13 TeV in
the $t-$ and $s-channel$ and in the tW associated mode.

The most recent measurement~\cite{t-ch} of single top production cross section at
14 TeV gives:
\begin{equation}
  \sigma_{t\mathrm{-ch.}} = 238\pm 13\mathrm{(stat)}\pm 29\mathrm{(syst)}\ \mathrm{pb}\ ,   
\end{equation}
and the ratio of top to antitop production cross setion is measured to be:
\begin{equation}
  R_{t\mathrm{-ch.}} = \frac{\sigma_{\mathrm{t}}}{\sigma_{\bar{\mathrm{t}}}} = 1.81 \pm 0.18\mathrm{(stat)} \pm 0.15\mathrm{(syst)}\ ,
\end{equation}
and is consistent with the asymmetry expected from the parton distribution functions (PDFs) of the proton.
The precision in the measurement of $R_{t\mathrm{-ch.}}$ does not yet allow
to set further constrints on proton PDF.
From single top production cross section, the CKM matrix element $\left|V_{\mathrm{tb}}\right|$ can be determined,
accounting for a possible presence of an anomalous Wtb coupling, which is taken into account with the form factor $f_{\mathrm{LV}}$:
\begin{equation}
  \left|f_{\mathrm{LV}}V_{\mathrm{tb}}\right| = 1.05\pm 0.07\mathrm{(exp)}\pm 0.02\mathrm{(th)}\ .
\end{equation}
All measurements agree so far with the Standard Model (SM) predictions.

\begin{figure}[ht]
% Use the relevant command for your figure-insertion program
% to insert the figure file.
\centering
\includegraphics[width=10cm,clip]{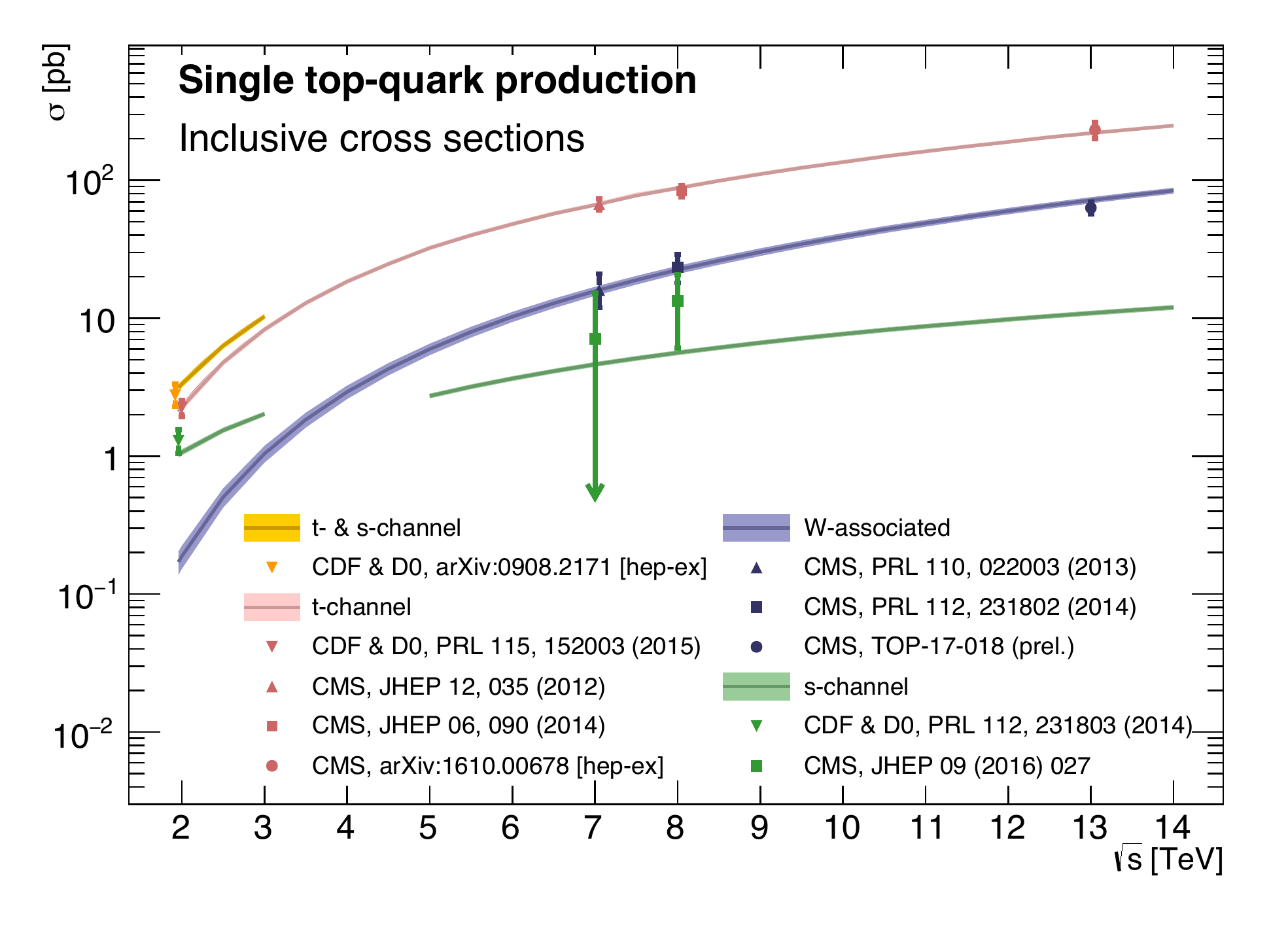}
\caption{Summary of single top cross section measurements by CMS, as function of centre-of-mass energy.
  This plot is from Ref.~\cite{TWiki}.}
\label{fig-1}       % Give a unique label
\end{figure}

A summary of single-top cross section measurements at different center-of-mass energies and in different channels by CMS is visible in Fig.~\ref{fig-1}.

\section{$\mathrm{t}\bar{\mathrm{t}}$ production}

The $\mathrm{t}\bar{\mathrm{t}}$ production has been measured at different center-of-mass energy and exploiting different top-quark decay channels.
One of the most recent measurements is in the $\ell$+jets channel at 13 TeV~\cite{ttbar}, which gives the following cross section:
\begin{equation}
  \sigma = 888\pm 2\mathrm{(stat)}^{+26}_{-28}\mathrm{(syst)}\pm 20\mathrm{(lumi)}\ \mathrm{pb}\ ,
\end{equation}
in agreement with the SM predictions.

The dependence of $\mathrm{t}\bar{\mathrm{t}}$ production cross section on the top quark pole mass allows to determine, indirectly:
\begin{equation}m_{\mathrm{t}}\mathrm{(pole)} = 170.6\pm 2.7\ \mathrm{GeV}\ ,
\end{equation}
which, though with less precision, is complementary to the direct mass measurements (see Sect.~\ref{sec:mt}).

A dedicated run at a reduced center-of-mass energy of 5~TeV in November 2015 allowed to measure
the $\mathrm{t}\bar{\mathrm{t}}$ production cross section at an energy intermediate between Tevatron and LHC~\cite{ttbar5tev}.
At lower energy the contribution of gluon fusion to the total production cross section lowers from 86\% at 13 TeV
to 73\%, allowing to set new constraints on the gluon PDFs, though with still a limitated precision due to the small
collected integrated luminosity of 27.4~pb$^{-1}$.

\begin{figure}[ht]
% Use the relevant command for your figure-insertion program
% to insert the figure file.
\centering
\includegraphics[width=10cm,clip]{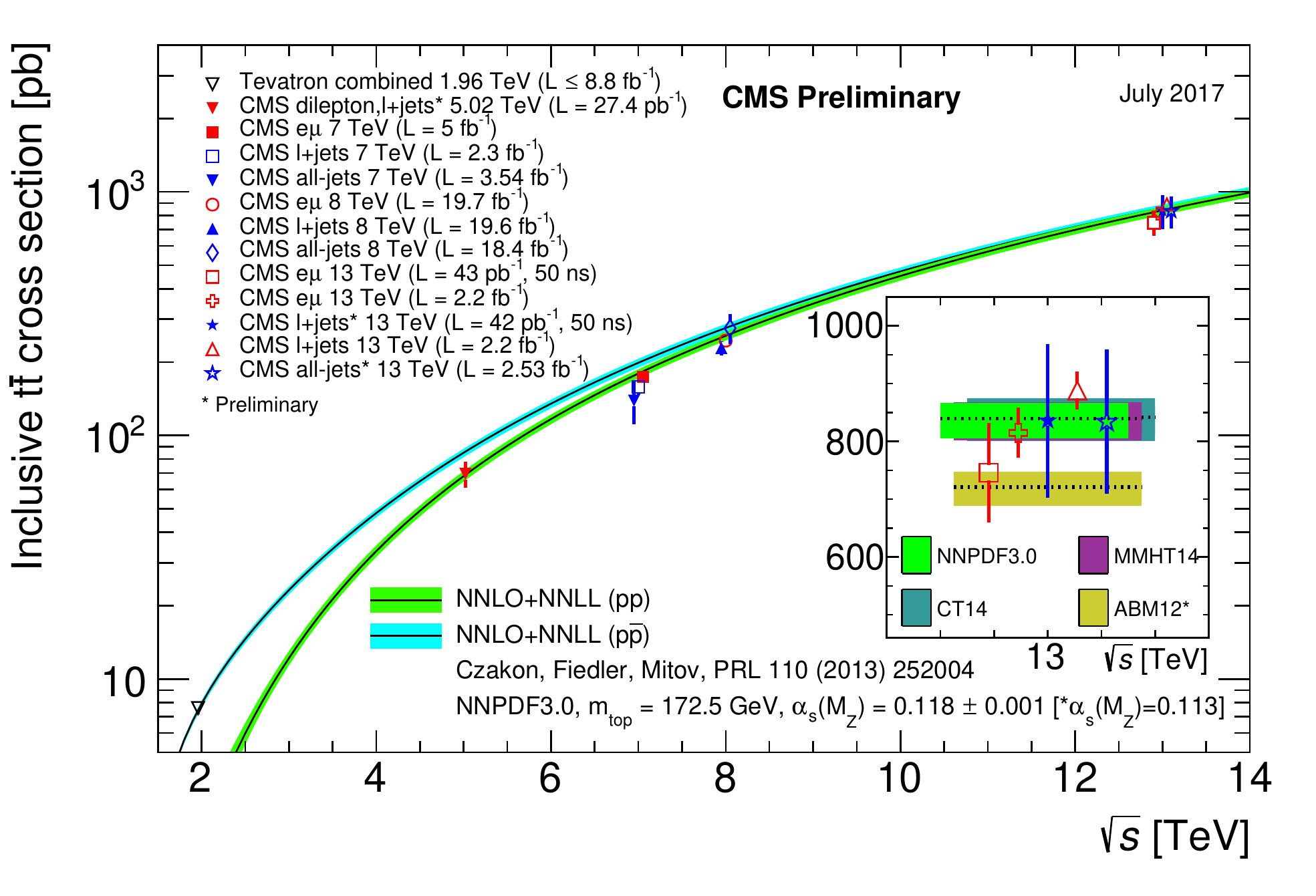}
\caption{Top quark pair cross section summary of most precise CMS measurements in the dilepton and $\ell$+jets channel in comparison with the theory calculation at NNLO+NNLL accuracy. The Tevatron measurements are also shown. This plot is from Ref.~\cite{TWiki}.}
\label{fig-2}       % Give a unique label
\end{figure}

A summary of the measured top-pair cross sections at different center-of-mass by CMS is visible in Fig.~\ref{fig-2}.

$\mathrm{t}\bar{\mathrm{t}}$ production has been studied in details with measurements of differential cross sections~\cite{ttbar_diff}
as a function of different kinematic variables with definition of the top quark at particle level. Measurements have been compared with SM prediction computed at the Next-to-leading order in QCD showing a good agreement with the theory.

In general, all  $\mathrm{t}\bar{\mathrm{t}}$ measurements performed at different center-of-mass energies show good agreement with theory and allow to set constraints on the proton PDFs.

\section{Top quark mass measurements}
\label{sec:mt}

The top quark mass is an important parameter in the SM, entering radiative corrections and affecting the stability of the Higgs potential.

CMS has measured $m_\mathrm{t}$ in different channels using different methods. The most precise measurement exploits the $\mu$+jets
channel at 13~TeV reconstructing the top mass with a kinematic fit that imposes the constaint of the W mass in the top decay products.
A requirement on the goodness-of-fit reduces spurious combinations. The top mass is fit together with a possible residual jet-energy
scale factor (JSF) in order to reduce systematic uncertainties due to jet-energy calibrations. The result is~\cite{mt}:
\begin{equation}
  m_{\mathrm{t}} = 172.62\pm 0.38\mathrm{(stat+JSF)}\pm 0.70\mathrm{(syst)}\ \mathrm{GeV}\ .
\end{equation}
Alternative approaches are also adopted for top mass measurements, in order to provide complementary
systematic uncertainties. One recent example is the measurement of $m_{\mathrm{t}}$ using spectra of kinematical
variables, such as the
invariant mass of the lepton and b quark from the top quark decay,
or the transverse invariant mass of the b-quark pair from the decay of the two top quarks~\cite{mtalt},
which exhibit a kinematic endpoint sensitive to $m_{\mathrm{t}}$. The measured top-quark mass value, combined with
different alternative approaches, is~\cite{mtatlcomb}:
\begin{equation}
  m_{\mathrm{t}} = 172.50\pm 0.21\mathrm{(stat)}\pm 0.72\mathrm{(syst)}\ \mathrm{GeV}\ .
\end{equation}
The result is in good agreement with the combination of `standard' techniques~\cite{mtcomb}:
\begin{equation}
  m_{\mathrm{t}} = 172.43\pm 0.13\mathrm{(stat)}\pm 0.46\mathrm{(syst)}\ \mathrm{GeV}\ ,
\end{equation}
but the achieved precision in the alternative procedures is not sufficient yet to improve the overall
combination, once combined with the `standard' measurements.

The top-quark mass has also been measured in highly boosted events
where the invariant mass spectrum has been used~\cite{mtjet} for jets reconstructed with a wide cone radius
parameter ($R$=1.2), aiming at the reconstruction of the decay product of a top quark into a single `merged' jet.
The uncertainty in the top quark mass measurement is so far limited ($\pm9$~GeV),
dominated by statistical uncertainty and jet energy scale uncertainty in the boosted regime.
Potential improvements and more constraints in systematic uncertainties are anyway possible in future
application of this method on largar data samples.

\begin{figure}[ht]
% Use the relevant command for your figure-insertion program
% to insert the figure file.
\centering
\includegraphics[width=8cm,clip]{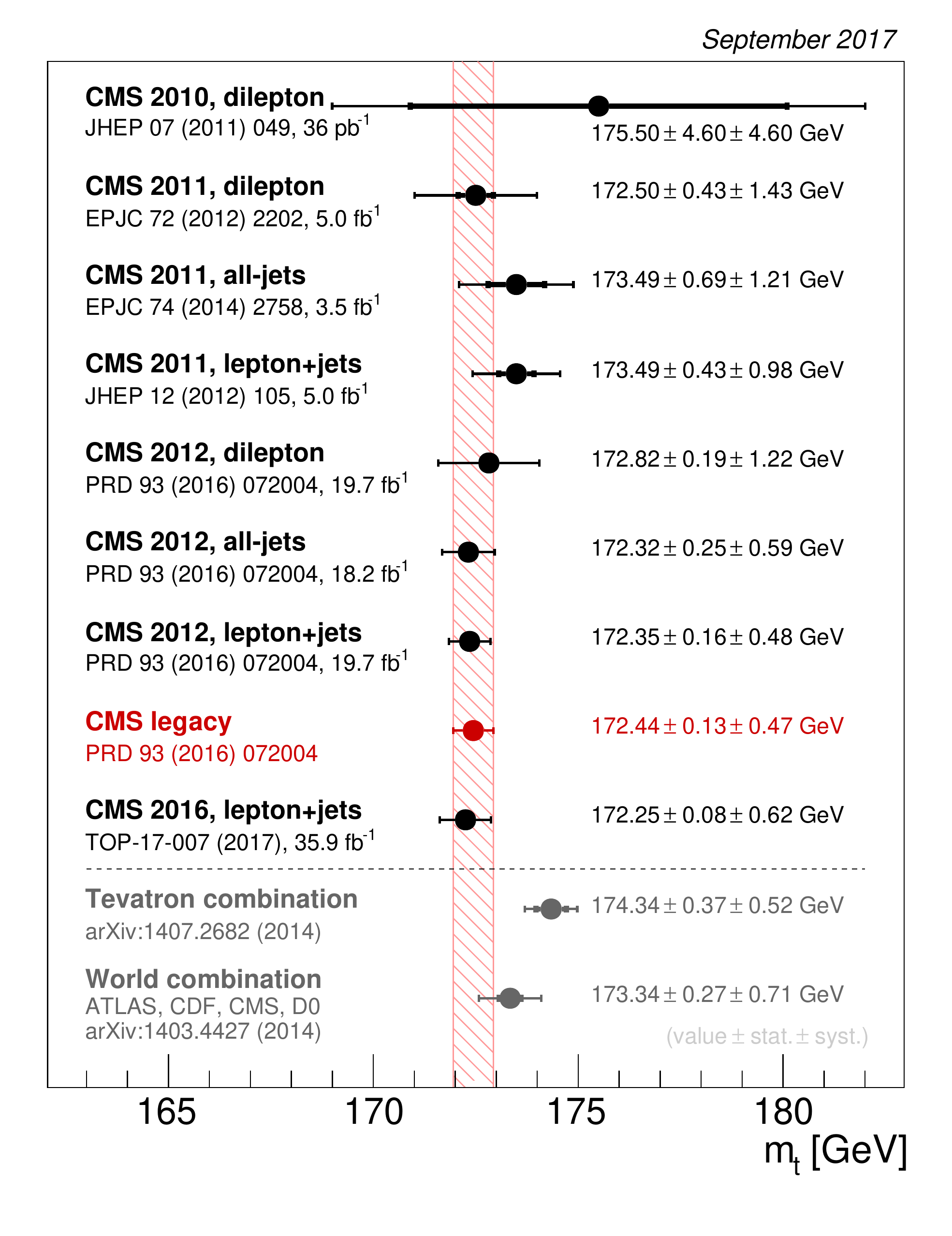}
\caption{Summary of Run-I CMS $m_{\mathrm{t}}$ measurements and their combination.
  This plot is from Ref.~\cite{mtfig}.}
\label{fig:mt}       % Give a unique label
\end{figure}

Figure~\ref{fig:mt} shows a summary of CMS $m_{\mathrm{t}}$ measurements done during Run-I.

\section{Top quark mass width}

The width of the top quark is predicted in the SM to be 1.35 GeV.
The invariant mass spectrum of the lepton and jet has been considered to measure
the top width~\cite{topw}, including boosted jet reconstruction, leading
to the following confidence interval, determined at the 95\% confidence level (CL):
\begin{equation}
0.6 \le \Gamma_{\mathrm{t}} \le 2.5\ \mathrm{GeV}\ .
\end{equation}
This measurement compares with the measurement of top width from indirect methods~\cite{topws},
which has about a 10\% uncertainty.

\section{Top pair production in association with vector bosons}

The production of a $\mathrm{t}\bar{\mathrm{t}}$ pair in association with a W or a Z
has been measured~\cite{ttWZ} using same-charge dilepton event with three- and four-lepton final states.
Jet and b-jet multiplicities have been used to enhance the signal-to-background ratio.
The following cross sections have been determined:
\begin{eqnarray}
  \sigma(\mathrm{pp}\rightarrow\mathrm{t}\bar{\mathrm{t}}\mathrm{Z}) & = & 1.00^{+0.09}_{-0.08}\mathrm{(stat)}^{+0.12}_{-0.10}\mathrm{(syst)}\ \mathrm{pb}\ ,\\
  \sigma(\mathrm{pp}\rightarrow\mathrm{t}\bar{\mathrm{t}}\mathrm{W}) & = & 0.80^{+0.12}_{-0.11}\mathrm{(stat)}^{+0.13}_{-0.12}\mathrm{(syst)}\ \mathrm{pb}\ ,\\
  \sigma(\mathrm{pp}\rightarrow\mathrm{t}\bar{\mathrm{t}}\mathrm{W}^+) & = & 0.58\pm 0.09\mathrm{(stat)}^{+0.09}_{-0.08}\mathrm{(syst)}\ \mathrm{pb}\ ,\\
  \sigma(\mathrm{pp}\rightarrow\mathrm{t}\bar{\mathrm{t}}\mathrm{W}^-) & = & 0.19\pm 0.07\mathrm{(stat)}\pm 0.06\mathrm{(syst)}\ \mathrm{pb}\ .\\
\end{eqnarray}

Constraints are set on Wilson coefficients of operators for ttZ and ttW production predicted by new physics models
beyond the SM (BSM). 

The production cross section of a top quark pair in association with a photon has been measured~\cite{ttg}
for semileptonic events in the fiducial volume defined applying the following requirement to the final-state particles:
\begin{itemize}
\item $\mathrm{e}$: $p_{\mathrm{T}} > 35\ \mathrm{GeV}$, $|\eta| < 2.5$,
\item $\mathrm{\mu}$: $p_{\mathrm{T}} > 26\ \mathrm{GeV}$, $|\eta| < 2.4$,
\item $\mathrm{\gamma}$: $p_{\mathrm{T}} > 25\ \mathrm{GeV}$, $|\eta| < 1.44$,
\item $\mathrm{\nu}$: $p_{\mathrm{T}}$ sum of the two neutrinos $> 20\ \mathrm{GeV}$.
\end{itemize}
Photon isolation is used to discriminate non-prompt photons.
The masured fiducial cross section, combining the electron and muon channels, relative to the top pair production
cross section, is:
\begin{equation}
  R = \sigma_{\mathrm{t}\bar{\mathrm{t}}\gamma}^{\mathrm{fid.}} / \sigma_{\mathrm{t}\bar{\mathrm{t}}} = (5.2\pm 1.1)\times 10^{-4}\ ,
\end{equation}
and the cross section, times the considered decay branching fraction is:
\begin{equation}
  \sigma_{\mathrm{t}\bar{\mathrm{t}}\gamma}\times {\cal B} = 515\pm 108\ \mathrm{fb}\ ,
\end{equation}
in agreement with the SM prediction of $591\pm 71\mathrm{(scale)}\pm 30\mathrm{(PDF)}\ \mathrm{fb}$.

\section{Single top and flavour-changing neutral current}

Events with a Z boson accompanying a single top quark allow to study anomalous flavour-changing neutral currents
foreseen in BSM models~\cite{fcnc} whose couplings can be modeled using effective operators.
Searches for a BSM signal have been done discriminating SM backgrounds using boosted decision trees (BDT)
multivariate discriminators in events with three leptons in the final state.
The observed yield is compatible with SM in different signal/control regions, and the
following cross section has been measured:
\begin{equation}
  \sigma(\mathrm{pp}\rightarrow\mathrm{tZq}\rightarrow\ell\nu\mathrm{b}\ell^+\ell^-\mathrm{q})
  = 10^{+8}_{-7}\ \mathrm{fb}\ ,
\end{equation}
to be compared with the SM prediction of 8.2 fb. The following limits have been determined
for the FCNC transition branching fractions at the 95\% confidence level (CL):
\begin{eqnarray}
{\cal B}(\mathrm{t}\rightarrow\mathrm{Zu}) & < & 0.022\%\ ,\\
{\cal B}(\mathrm{t}\rightarrow\mathrm{Zc}) & < & 0.049\%\ .
\end{eqnarray}
BSM models predict FCNC branching fractions up to $\sim 10^{-4}\div 10^{-3}$.

\section{Four top production}
An example of a process that is very rare in the SM, but may be significantly enhanced in BSM
models, is the production of four top quarks.
The SM production cross section prediction is only 1.3 fb.
Single lepton and opposite-sign dilepton + jets events have been studied with
an extensive usage of BDT techniques~\cite{ftop}, combining  global event and jet properties.
A first BDT ranks 3-jet combinations consistent with all-hadronic top quark decays based on invariant masses and b-tag information;
then, a second BDT uses the ranking from the first BDT plus global event kinematic variables to identify signal events.
The upper limit to four-top production cross section is, at the 95\% CL:
\begin{equation}
  \sigma_{\mathrm{t}\bar{\mathrm{t}}\mathrm{t}\bar{\mathrm{t}}} < 94\  \mathrm{fb} = 10.2\times \mathrm{SM}\ .
\end{equation}
Combined with the same-sign dilepton channel, the limit improves to:
\begin{equation}
  \sigma_{\mathrm{t}\bar{\mathrm{t}}\mathrm{t}\bar{\mathrm{t}}} < 69\  \mathrm{fb} = 7.4\times \mathrm{SM}\ .
\end{equation}

\section{Future perspectives}

Long-term perspectives studies were prepared for  the 2014 Snowmass meeting~\cite{fp1}
and for ECFA 2016~\cite{fp2} based on extrapolations of the LHC integratedluminosity and potential analysis technique improvements.

Prospects for Yukawa coupling measured with ttH Higgs boson production, studied in different final states, give
the following possible precision levels:
\begin{eqnarray}
  \kappa_{\mathrm{t}} & \rightarrow & 14\div 15\%\ (300\ \mathrm{fb}^{-1})\ , \\
  \kappa_{\mathrm{t}} & \rightarrow & 7\div 10\%\ (3\ \mathrm{ab}^{-1})\ . 
  \end{eqnarray}
Expectations for FCNC, still based on old preliminary results at 8 TeV, give the following
potential upper limits:
\begin{eqnarray}
{\cal B}(\mathrm{t}\rightarrow\mathrm{Zu},\,\mathrm{c}) & < 1\times 10^{-4}\ (3\ \mathrm{ab}^{-1})\ ,\\
{\cal B}(\mathrm{t}\rightarrow\gamma\mathrm{u}) & < 2.6\times 10^{-4}\ (3\ \mathrm{ab}^{-1})\ .
\end{eqnarray}

The top-quark mass may be measured with a precision down to about 200 MeV, exploiting a combination
of standard and alternative techniques.

\section{Conclusions}

Many precision measurements have been performed by CMS at 13 TeV and more are ongoing after legacy measurements were established at 7 and 8 TeV.
No deviation from the Standard Model have been observed so far, but the search continues with more and more precision exploiting the high luminosity of future LHC runs.
More channels are opening with higher luminosity, with the possibility to probe possible deviations from the Standard Model.

\end{document}